\documentclass{article}
\usepackage[T1]{fontenc} 
\usepackage[utf8]{inputenc} 
\usepackage{ismir,amsmath,cite,url}
\usepackage{graphicx}
\usepackage{color}
\usepackage{amsmath}
\usepackage{comment}
\usepackage{subcaption}
\usepackage{bm}
\usepackage{caption}
\usepackage{amssymb}
\usepackage{calc}
\usepackage{lineno}

\title{Learning multidimensional disentangled representations of instrumental sounds for musical similarity assessment}

\multauthor
{Yuka Hashizume$^1$ \hspace{1cm} Li Li$^1$ \hspace{1cm} Atsushi Miyashita$^1$} { \bfseries{Tomoki Toda$^1$ \hspace{1cm}}\\
  $^1$ Nagoya University , Japan\\
{\tt\small hashizume.yuuka@g.sp.m.is.nagoya-u.ac.jp,}\\ 
{\tt\small li.li@g.sp.m.is.nagoya-u.ac.jp,}\\
{\tt\small miyashita.atsushi@g.sp.m.is.nagoya-u.ac.jp,}\\ {\tt\small tomoki@icts.nagoya-u.ac.jp}
}

\def\authorname{Y. Hashizume, L. Li, A. Miyashita, and T. Toda}

\makeatletter
\def\bstctlcite{\@ifnextchar[{\@bstctlcite}{\@bstctlcite[@auxout]}}
\def\@bstctlcite[#1]#2{\@bsphack
\@for\@citeb:=#2\do{%
\edef\@citeb{\expandafter\@firstofone\@citeb}%
\if@filesw\immediate\write\csname #1\endcsname{\string\citation{\@citeb}}\fi}%
\@esphack}
\makeatother

\begin{document}
\bstctlcite{IEEEexample:BSTcontrol}

\maketitle
\begin{abstract}
To achieve a flexible recommendation and retrieval system, it is desirable to calculate music similarity by focusing on multiple partial elements of musical pieces and allowing the users to select the element they want to focus on. A previous study proposed using multiple individual networks for calculating music similarity based on each instrumental sound, but it is impractical to use each signal as a query in search systems. Using separated instrumental sounds alternatively resulted in less accuracy due to artifacts. In this paper, we propose a method to compute similarities focusing on each instrumental sound with a single network that takes mixed sounds as input instead of individual instrumental sounds. Specifically, we design a single similarity embedding space with disentangled dimensions for each instrument, extracted by Conditional Similarity Networks, which is trained by the triplet loss using masks. Experimental results have shown that (1) the proposed method can obtain more accurate feature representation than using individual networks using separated sounds as input, (2) each sub-embedding space can hold the characteristics of the corresponding instrument, and (3) the selection of similar musical pieces focusing on each instrumental sound by the proposed method can obtain human consent, especially in drums and guitar.
\end{abstract}
\section{Introduction}
\label{intro}
The music market is proliferating~\cite{ifpi}, where the number of musical pieces available on music streaming services is about 100 million today~\cite{apple}. 
Subsequently,
it is impossible for users to listen to all of them to find their favorite music. Therefore, Music Information Retrieval (MIR) technologies, such as music recommendation systems, are needed to help users find their favorite music efficiently.
\par 
One effective method for MIR is to define similarities between musical pieces and use the user's favorite piece as a query for searching or recommendation. When using this method, it is essential to design suitable similarity criteria for calculating music similarity. 
With the advent of deep learning, data-driven feature extraction has shown to be effective in improving the performance of MIR systems~\cite{Hamel2010, Elbir2020}. For example, methods have been proposed to learn feature representation by metric learning using tags and labels~\cite{park2018, Clevelan2020, Lu2017}.
These methods calculate music similarity by evaluating an entire musical piece, a mixture of various sounds, using a single criterion. However, music is a complex structure with various significant elements, and what users focus on when listening to music varies from user to user. To achieve a more flexible MIR system, it is desirable to calculate music similarity by focusing on multiple partial elements of musical pieces and allowing the users to select the element they want to focus on.
\par
The previous 
studies~\cite{has2022, has2022au} proposed a music similarity calculation method focusing on individual instrumental sounds in a musical piece, where networks were trained for each instrument using single instrumental signals. 
One limitation of this method is the need of individual instrumental sounds
not only in training but also in inference, 
where individual instrumental sounds are difficult to obtain in practice.
To address this issue, the use of instrumental sounds separated from mixed sounds was also investigated. 
However, using the separated sounds often suffering from artifacts resulted in lower accuracy than using the original instrumental sounds. 
\par
Another promising approach is to extract feature representation for each instrument directly from the mixed musical piece.
Disentanglement feature representation learning 
is one such method that can extract
several different conceptual representations from a single input~\cite{Bengio2013}
, which has been adopted to disentangle the speaker identity and noise in the speech domain~\cite{Hsu2018, Hsu2019}, and timbre and pitch information in the music domain\cite{Hung2018, Luo2019, tanaka21}.
Veit et al.~\cite{CSN} proposed Conditional Similarity Networks (CSNs) that learns embeddings differentiated into semantically distinct subspaces that capture the different notions of similarities in the image domain. Lee et al.~\cite{Disen} applied CSNs to the music domain and designed an embedding space such that each subspace represents the four similarity metrics, genre, mood, instrumentation, and tempo.
\par
In this paper, we propose a method to compute similarities focusing on individual instrumental sounds using mixed sounds as input in one network. 
The proposed network is trained with metric learning to embed mixed musical pieces into differentiated feature space, where each subspace selected by a binary mask represents musical characteristics when focusing on a particular instrument.
To successfully train the network, we implement new ideas for the training, such as the use of pseudo-mixed pieces, an auxiliary loss, and pre-training.
In the experiments, we investigate whether more accurate feature representations can be obtained than using conventional methods, whether each subspace holds the characteristics of the assigned instrument sounds, and whether the learned similarity criterion matches human perception.

\section{Related Research}
\label{relate}
\subsection{Music similarity calculation of individual instrumental sounds using metric learning}
\label{individual}
In deep metric learning with a triplet loss~\cite{Ailon2015}, a distance metric is trained with a triplet of samples, where one is considered as an anchor, and the other two are considered as positive and negative samples. Here, the positive sample should be more similar to the anchor than the negative one. Lee et al.~\cite{Disen} proposed the track-based music similarity; segments from the same track as the anchor are defined as positive samples, and those from different tracks from the anchor are defined as negative samples. They used the music signal itself as input.
\par
To achieve a highly flexible MIR system, the previous 
studies
proposed a music similarity calculation method focused on each individual instrumental sound~\cite{has2022, has2022au}. In this method, metric learning with triplet loss was applied to individual instrumental sources such as drums, bass, piano, and guitar. Positive and negative samples were defined by track-based similarity. Different networks were separately trained for individual instrumental sounds. Consider the case where the instrumental sounds included in the musical piece users want to query for are not available when using the search, separated instrument signals, which were extracted from the mixed music signals using an instrumental source separation method, were applied to the method.
\par
Letting $x_i^{(a)}$, $x_i^{(p)}$, and $x_i^{(n)}$ denote the $i$-th anchor, positive sample, and negative sample, respectively, the triplet $t_i$ is constructed as a set of $\{x_i^{(a)}, x_i^{(p)}, x_i^{(n)}\}$, where $i = 1, \ldots, I$ denotes the index of training samples. The triplet loss is defined as
\begin{equation}
    \mathcal{L}(t_i)=\max\{d(x^{(a)}_i,x^{(p)}_i)-d(x^{(a)}_i,x^{(n)}_i)+\Delta,0\},
\end{equation}
where $d$ is a distance function for measuring the distance between two audio samples, such as the Euclidean distance, and $\Delta$ is a margin value, which defines the minimum distance between the positive and negative samples.

\subsection{Conditional Similarity Network}
\label{CSN}
To measure the similarity between images considering multiple notions of similarity, Veit et al. \cite{CSN} proposed Conditional Similarity Networks (CSNs) that learns embeddings differentiated into semantically distinct subspaces that capture the different notions of similarities.

The notions of similarity are, for example, the height of the shoes’ heels, etc., in the example where the input is an image of a shoe.
In this method, a network extracting an embedding representation is learned by the triplet loss using masks. For the triplet loss, samples $x^{(a)}$, $x^{(p)}$ and $x^{(n)}$ are selected according to the condition $c$ that is defined as a certain notion of similarity. Namely, in notion corresponding to the condition $c$, $x^{(p)}$ is more like $x^{(a)}$ than $x^{(n)}$.
To disentangle the embedding space, a mask is applied to all dimensions except the dimension corresponding to the notion to be considered, in the triplet loss calculation. The network is given by function $f(\cdot)$, and $m_c$ is a mask that activates only the dimension corresponding to the condition $c$. The masked distance function between two images $x_i$ and $x_j$ is given by
\begin {align}
\label{CSNtriplet}
D(x_i,x_j;m_c)=\parallel f(x_i)m_c-f(x_j)m_c\parallel_2.
\end{align}
Thus, the triplet loss can be written as follows.
\begin {align}
 &\mathcal{L}_T(x^{(a)},x^{(p)},x^{(n)},c;m)=\notag \\
 &\max\{0,D(x^{(a)},x^{(p)};m_c)-D(x^{(a)},x^{(n)};m_c)+\delta\}.
\end{align}
\par
Lee et al. \cite{Disen} also proposed the disentangled multidimensional metric learning for music similarity using CSNs. They used musical genre, mood, instrument, and tempo for the notions of similarity. 

\begin{figure}[t]
\centering
\scalebox{0.25}{
\centerline{\includegraphics{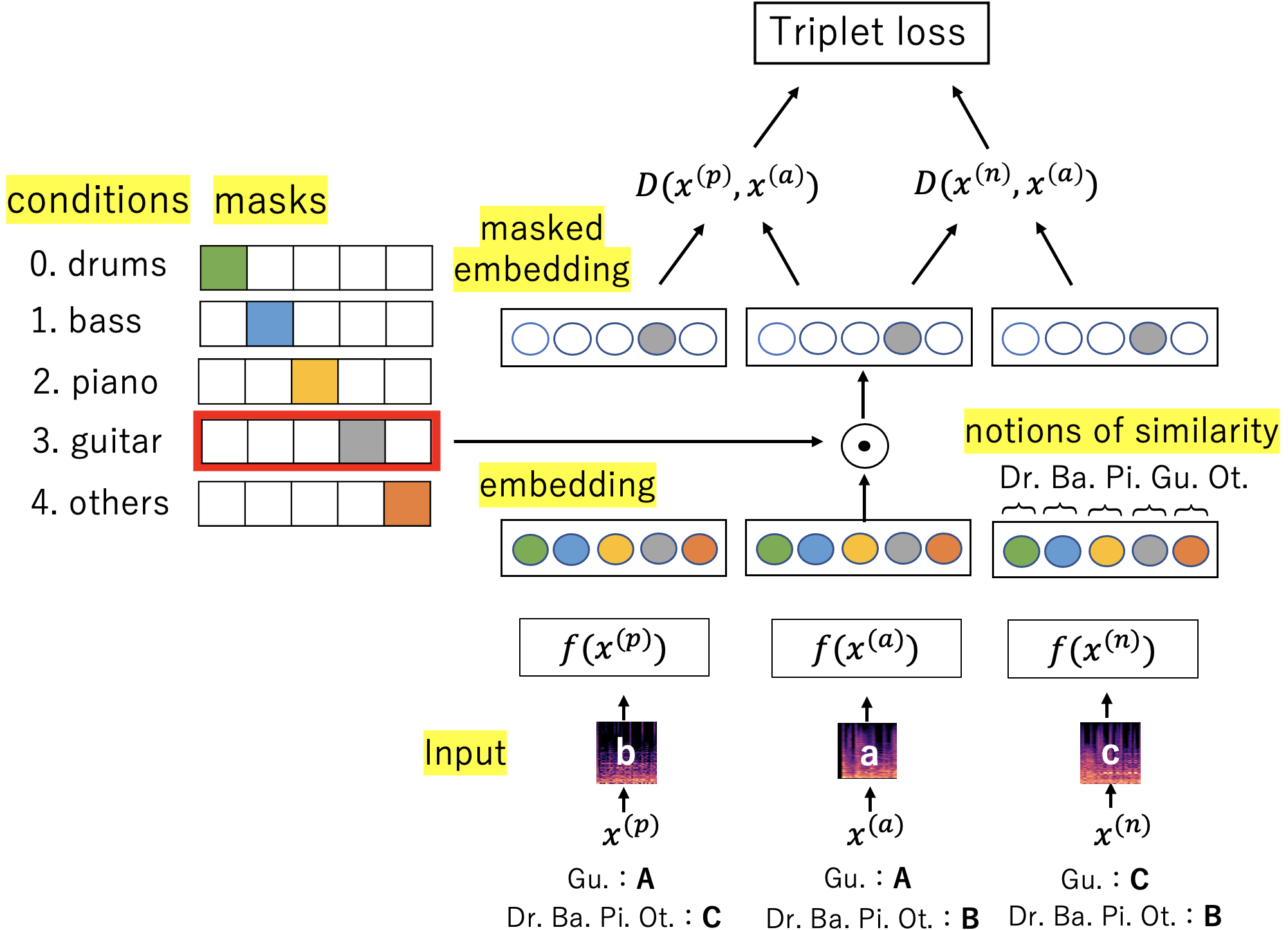}}}
\caption{Overview of the proposed method.  The $x^{(a)}$, $x^{(p)}$, and  $x^{(n)}$ denote the anchor, positive, and negative samples, respectively. The ‘Dr.,' ‘Ba.,' ‘Pi,' ‘Gu.' and ‘Ot.' are drums, bass, piano, guitar, and others, respectively. The $\boldsymbol{A}$, $\boldsymbol{B}$, and $\boldsymbol{C}$ indicate the ID of the musical piece in which each instrumental sound is originally included. This figure shows an example of setting the condition to guitar, where an anchor sample "a" and a positive sample "b" are extracted respectively from two pseudo-mixed pieces $A^{(gu)}_B$ and $A^{(gu)}_C$containing different segments of the guitar sounds of the same piece A. From each sample, the embedded representation is extracted by the network and is masked so that the dimension to which the guitar is assigned only validates in the triplet loss calculation.}
\label{prop}
\end{figure}

\begin{figure}[t]
\centering
\scalebox{0.4}{
\centerline{\includegraphics{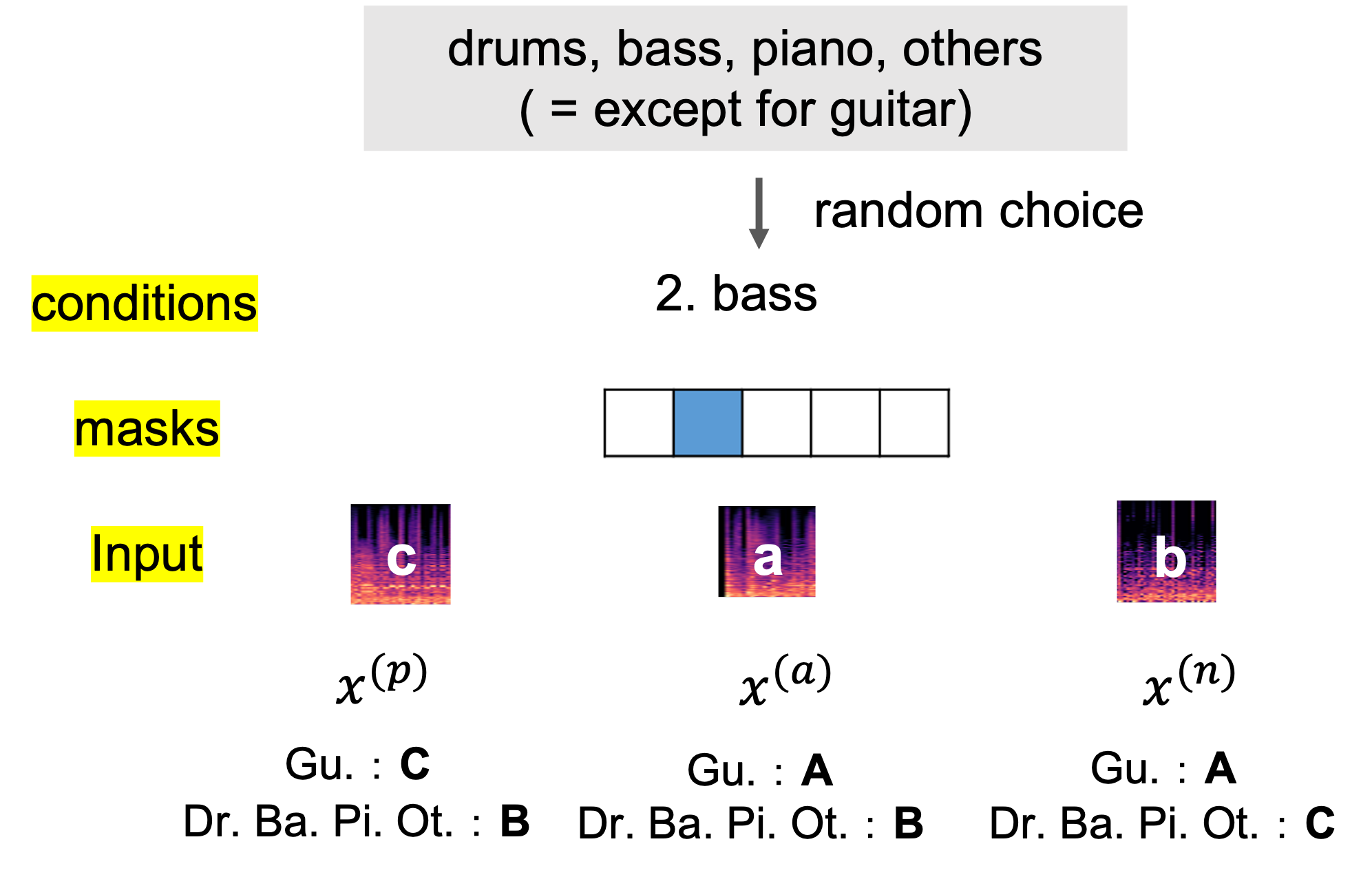}}}
\caption{Interchanged triplet to be used in addition to the basic triplet. The negative sample "c" in Fig. 1 is used as a positive sample, and the positive sample "b" in Fig.~\ref{prop} is used as a negative sample by setting the condition to another except for guitar, e.g., bass.}
\label{trip2}
\end{figure}

\begin{figure}[t]
\centering
\scalebox{0.35}{
\centerline{\includegraphics{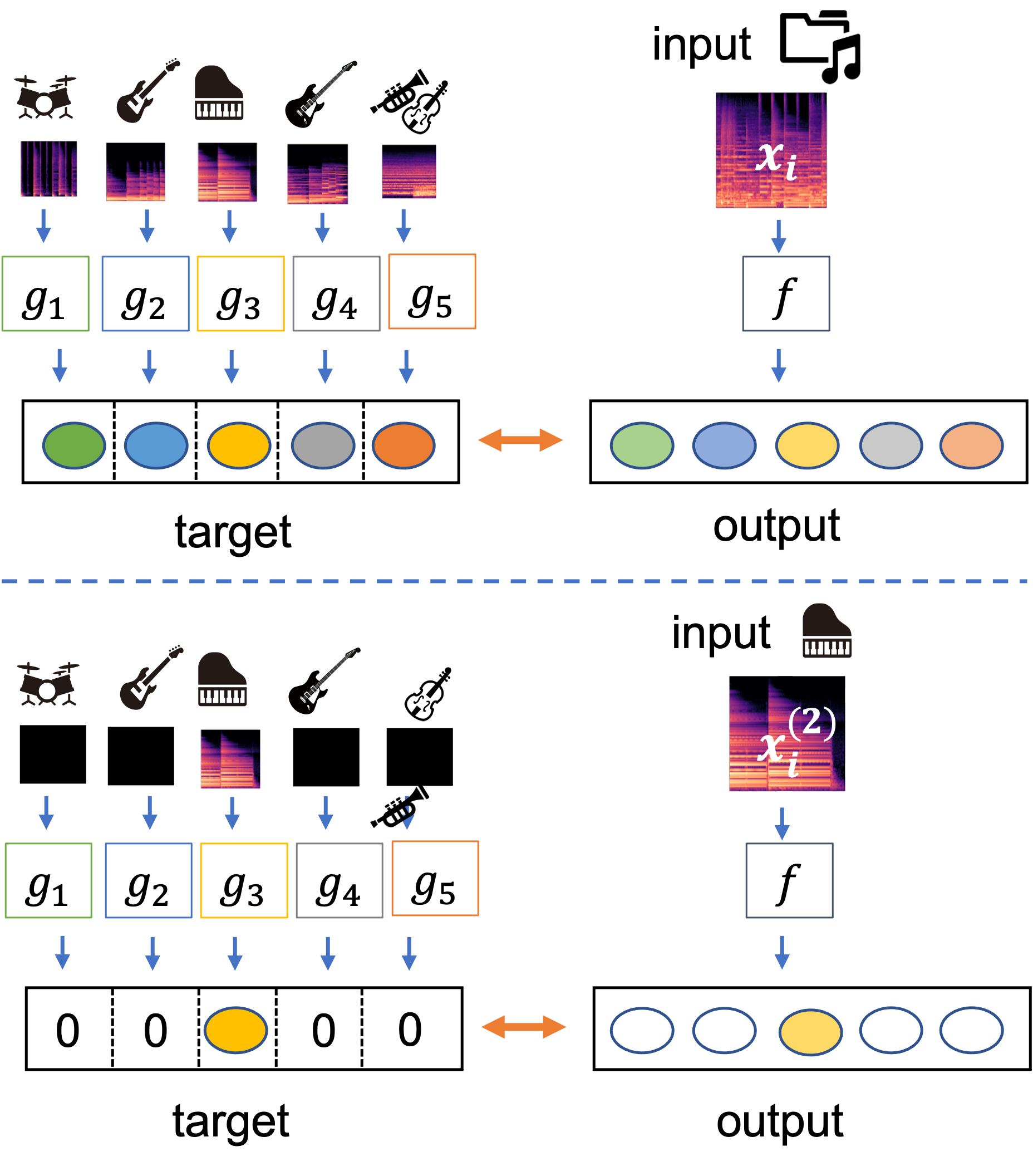}}}
\caption{Generation of target embeddings for auxiliary loss calculation. The upper figure shows the target embedding generation for the mixed sounds. The lower figure shows that for each of the individual instrumental sounds, which is used for the pretraining only.}
\label{pretrain}
\end{figure}

\section{Proposed method}
\label{prop-method}
\subsection{Triplet loss with mask}
\label{triplet}
In this study, the CSNs described in Section~\ref{CSN} are used, with the notion of similarity defined as whether or not musical pieces are similar to each other, focusing on individual instrument sounds.
\subsubsection{Pseudo mixed sound}
\label{pseudo}
When training CSNs, each subspace is trained with triplets sampled by using different labels based on each similarity notion~\cite{CSN}. However, there is no label that evaluates whether or not musical pieces are similar to each other focusing on individual instrument sounds. Although the track information was used in the previous study~\cite{has2022}, we found that using it in our proposed framework leads to all subspaces being trained on the same criteria. 
\par
To successfully train the disentangled subspaces, in this study, we propose a method to create a pseudo-musical piece by mixing instrumental sounds in different musical pieces and using them as input. Specifically, when the drum sound contained in a piece A is called drum sound A, a pseudo-mixed piece $A^{(dr)}_B$ can be created by mixing the drum sound A with other instrumental sounds from another piece B. In this case, we can say that segment 1, randomly extracted from the original piece A (i.e., $A^{(dr)}_{A}$), and segment 2, randomly extracted from the pseudo-mixed piece $A^{d}_{B}$, are similar in drum sounds but not similar in the sounds of other instruments. On the other hand, the segments extracted from the pseudo-mixed pieces $A^{(dr)}_C$ and $B^{(dr)}_C$ are not similar to each other in drums, but they are similar in instrumental sounds other than drum sounds.

\subsubsection{Loss function formulation}
\label{lossfunc}
We divide the mixed sound into three seconds segments $x$ and input them into the network to train and obtain $f(x)$.
We design an embedding representation space whose dimensions are disentangled by the five instruments and define $c$ as the condition where $c=0, 1, 2, 3, 4$ represent drums, bass, piano, guitar, and others, respectively. Let $D$ be the number of dimensions of subspace assigned for one instrument, and the subspace assigned to condition $c$ are as $f[cD:(c+1)D-1]$. The following formula defines the $m_c$ as a mask that leaves the subspace corresponding to each condition $c$ and sets the other dimensions to 0. The triplet loss in CSNs shown in Equation~\ref{CSNtriplet} is used for learning using a mask as follows.
\begin {align}
m_c[d] = \left\{
\begin{array}{ll}
1, & (cD\leqq d<(c+1)D)\\
0, & (\mbox{otherwise}).
\end{array}
\right.
\end{align}

For example, when $c=0$, i.e., drums, the triplet is chosen from the pseudo-mixed pieces described in Section~\ref{pseudo} as $\{A^{(dr)}_B, A^{(dr)}_C, C^{(dr)}_B\}$. An overview of the proposed method is shown in Fig.~\ref{prop}. The triplets extracted in this way are called the basic triplet. We further add triplets of interchanged positive and negative samples under different condition $c$, in addition to the basic triplet, to allow each subspace to learn a different similarity criterion explicitly. This additional triplet extraction method is shown in Fig.~\ref{trip2}. Note that the basic triplet’s negative sample is selected so there is no conflict between the basic triplet and the additional triplet.

\subsection{Auxiliary loss using individual network embeddings}
\label{mse}
To encourage each subspace to represent the characteristics of each instrumental sound, we use the following auxiliary loss function with the embedding $g_c( \cdot)$ extracted from each of the networks trained separately with individual instrumental sounds in the previous study~\cite{has2022, has2022au}, where $g_c( \cdot),  (c=0,1,2,3,4)$ represents the embeddings corresponding to drums, bass, piano, guitar, and others. In order to ensure that each of the individual instrumental sounds does not affect dimensions other than the corresponding dimension, the embedding is set to the zero vector if the input sample doesn't contain the corresponding instrumental sound. The $x^{(c)}_i$ is the $i$-th individual instrumental sound segment corresponding to condition $c$, contained in the $i$-th mixed sound segment $x_i$. The $\frac{y_i}{\parallel y_i \parallel_2}$ is the target embedding for the network training, which is created by concatenating embedding features extracted from $x^{(c)}_i$ using the individual networks and divided by the norm. Figure~\ref{pretrain} shows a way of creating the target embeddings. As shown in the lower part of the figure, the individual instrumental sound $x^{(c)}_i$ included in the mixed sound $x_i$ is also used to generate the target embeddings, which are helpful in the pre-training stage, as explained later in Section~\ref{pseknn}. The formulations of the auxiliary loss function $\mathcal{L}_M$ and the target embedding are as follows:

\begin {align}
&\mathcal{L}_M=\left|\left| f(x_i) - \frac{y_i}{\parallel y_i \parallel_2} \right|\right|_2,\notag
\\&y_i[d] = g_c(x^{(c)}_i),  (cD\leqq d<(c+1)D).
\end{align}

\subsection{Training network}
\label{training}
The final loss function $\mathcal{L}$ is as follows, where $\lambda$ is the hyperparameter that weights the loss function $\mathcal{L}_M$.
\begin {align}
\mathcal{L}=\mathcal{L}_{T}+\lambda \mathcal{L}_M
\end{align}
In this study, $\mathcal{L}_M$ is also used for pre-training.

\begin{figure}[t]
\centering
\scalebox{0.52}{
\centerline{\includegraphics{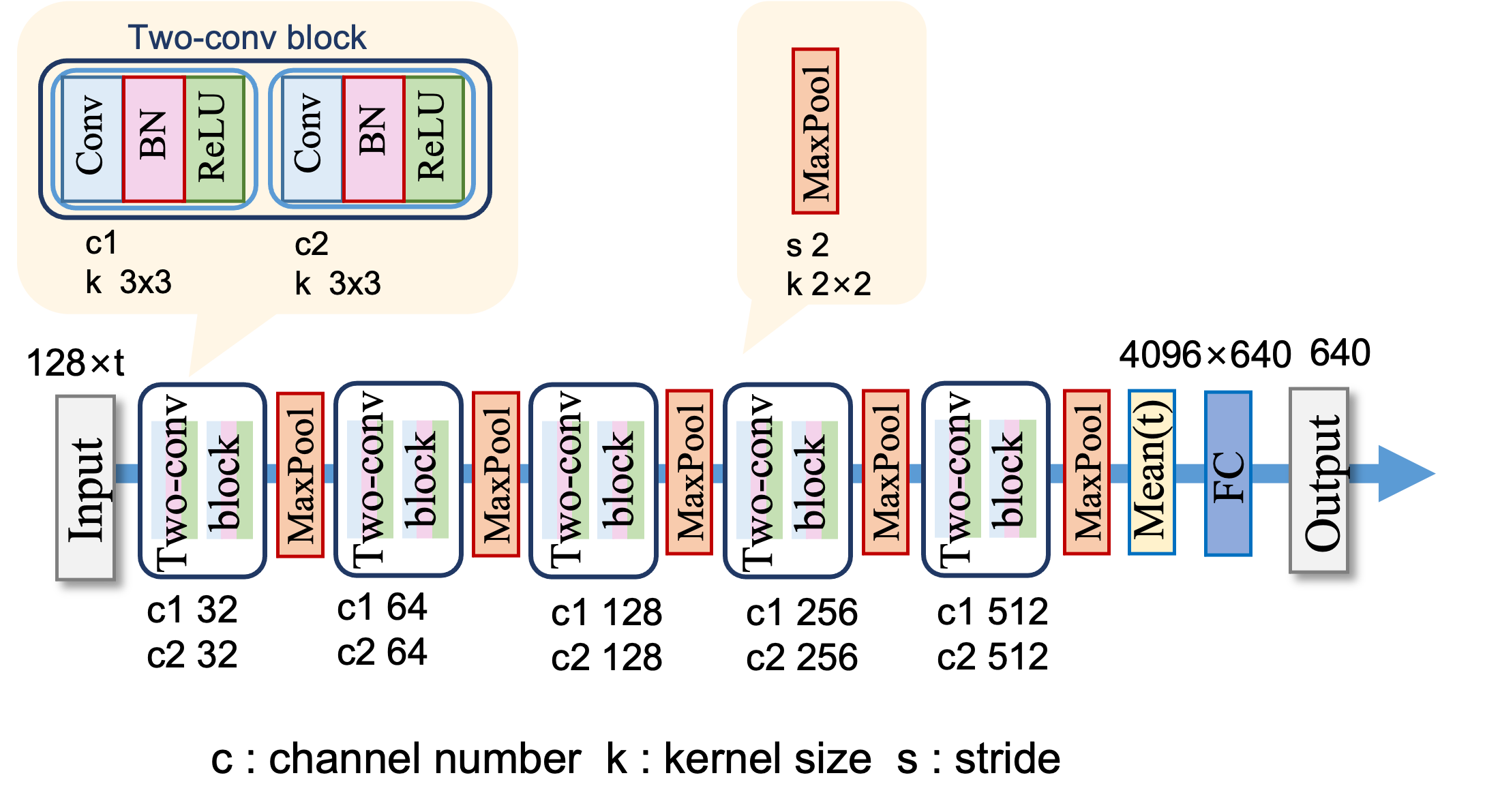}}}
\caption{Network architecture of the network. The ‘c,' ‘k,' and ‘s' denote the channel number, kernel size, and stride. “Conv," and “FC" denote the convolutional and fully connected layers, respectively. “BN” means batch normalization. “Mean(t)” means to take an average in the time direction. The numbers above input, output, and “FC" are their sizes.}
\label{network}
\end{figure}

\section{Experimental evaluation}
\subsection{Experimental conditions}
\subsubsection{Dataset}
The dataset we used is slakh \cite{slakh}, which contains non-vocal musical pieces and their stems. Following slakh's recipe, individual instrumental sounds, drums, bass, piano, and guitar sound, were created from their stems, and the stems that did not fit into any of the four instruments were mixed as “others.” 
\par
We used these musical pieces and instrumental sounds for training with $\mathcal{L}_M$. In addition, the pseudo-mixed pieces mentioned in Sectin~\ref{pseudo} were created for training with $\mathcal{L}_T$. To preserve the music-like nature of the music, the data set was classified by tempo, and the instrumental sounds contained in musical pieces belonging to the same tempo group were allowed to be mixed together. When mixing, the onsets were aligned with the onset of the musical pieces to which each instrumental sound belonged. Under these rules, multiple different pseudo-mix pieces containing the same instrumental sound were generated.
\subsubsection{Network}
The network shown in Fig.~\ref{network} similar to the encoder portion of U-Net~\cite{unet}, used in a previous study on singing voice separation~\cite{singunet}, was trained to extract a 640-dimensional embedding vector from a mel-spectrograms as embedding representations. 
The 640-dimensional embedding representation was aimed to have 128-dimensional subspaces assigned to each of the five instruments. 

\subsubsection{Pretraining}
We used the original instrumental sounds of 200 pieces from the slakh train set to train the individual networks by the conventional method~\cite{has2022, has2022au} to extract each instrument's embedding representation for the target of $\mathcal{L}_M$. Each musical piece for training was split into three-second segments with 50\% overlap, and the first 40 segments were used in each musical piece, excluding the silent parts. Fully Convolutional Networks (FCNs) used in the previous study~\cite{has2022au} were used for learning individual networks. 
\par
Then, we pre-trained the network shown in Fig.~\ref{network} using the outputs of individual networks trained following the above procedure. The inputs were segments from 200 musical pieces, and each target was obtained from instrumental sound segments contained in each input. 

\subsubsection{Training}
In the same manner as the pretraining, the 1200 data from the slakh train set were divided into three-second segments with overlap. Then, the network was trained by the triplet loss using pseudo-mixed pieces. The anchors were randomly selected from these segments, and 20,000 triplets were created for metric learning by triplet loss. The margin of the triplet loss function was set to 0.2.
\par
The auxiliary loss function described in Section~\ref{mse} was also used simultaneously, where the auxiliary loss was calculated for only the anchors. The weighting parameter between two losses $\lambda$ was set to 0.1. The batch size was set to 64. The number of epochs was set to 400.

\subsubsection{Testing}
The data used for the test consisted of 136 pieces of 151 slakh test data, excluding pieces with less than 10\% of segments when the instrumental sounds were divided, eliminating the silent parts. All segments were converted to mel-spectrograms, normalized, and used as input for the training and test.

\subsection{Evaluation method}
We conducted experimental evaluations to investigate whether the following two purposes of this study were achieved, (P1) to learn an embedding representation in which similar pieces are close and dissimilar pieces are far from each other and (P2) to output the similarity focusing on each instrumental sound in the subspace assigned to each instrument.
\subsubsection{Accuracy of embedding}
\label{knn}
In the evaluation on P1, we used the accuracy of music IDs predicted using close samples' representation as used in the previous study~\cite{has2022, has2022au}. This evaluation was based on the assumption that instrumental sounds that consist of different time segments of the same musical piece should be more similar than those of different musical pieces. Subjective evaluation experiments later confirm whether this assumption fits the human senses.
Specifically, we used the K-nearest neighbor (kNN) method to predict the music IDs of the test segments. The music IDs of all test segments except the one to be predicted were assumed to be known. We embedded all test segments into the learned representation space and predicted the music ID of each test segment by a majority vote using the IDs of the top five nearest test segments. To check whether each subspace has learned the similarity criterion, the above evaluation was performed for each subspace by applying a mask that enables only each subspace, as was done when learning using triplet loss.

\subsubsection{Representation of each subspace}
\label{pseknn}
In the evaluation on P2, the pseudo-mixed pieces were created for the test pieces using the same method as described in Section~\ref{pseudo}. When the subspace was used to measure the similarity focusing on the corresponding instrument, we evaluated whether the distance between the pseudo-mixed samples containing the focused instrumental segments from the same musical piece is smaller than the others, using the same kNN-based method as described in Section~\ref{knn}.
\par
It is possible that the subspace for the focused instrumental sound is affected by the other instrumental sounds. If we use the pseudo-mixed segments containing the other instrumental segments extracted from the same musical piece as in the pseudo-mixed segment to be inferred in the kNN-based method, it is possible that the other instrumental segments will affect the inference result. To avoid this issue, we removed all of the possible pseudo-mixed segments causing this issue, i.e., when a pseudo-mixed segment from $A^{(dr)}_B$ was inferred in the evaluation of the subspace on the drums, we removed all of the pseudo-mixed segments from $A^{(dr)}_B$ from the k-NN dataset. Ten musical pieces were randomly selected from the test pieces for this evaluation. For each of the ten musical pieces, four pseudo-mixed pieces were created in the method explained in Section~\ref{pseudo}, and they were divided into 10 seconds segments. In other words, the segments from 40 pseudo-mixed pieces with ten music IDs were used.
\subsubsection{Subjective evaluation}
\label{sub}
Subjective evaluation experiments were conducted to confirm whether each subspace represented a similarity criterion such that the distance between sounds was small enough that humans would perceive them to be similar when listening to each assigned instrumental sound. The procedure was described below, using the evaluation of drum sounds as an example.
We randomly selected eight musical pieces $X$ from the test pieces, clipped 10 seconds of non-silence from the drum sounds that consist of $X$, and denoted them as $x$. From the different pieces from $X$, we randomly selected pieces $A$, $B$, and $C$, cut out 10 seconds of each sound sample in the same way, and denoted them as $a$, $b$, and $c$, respectively. Also, let $y$ be a 10-second sound sample extracted from a different time part from $x$ in piece $X$. For these sound samples, two sound sets $\{x, a, b\}$ and $\{x, y, c\}$ were created. This process was repeated for bass, piano, and guitar, and the 64 sets created in this way were played randomly to the subjects, who were asked whether $a$ or $b$ sounded more like $x$ or whether $y$ or $c$ sounded more like $x$.
If the subject selected a sample with a smaller distance from the drum sound $x$ in the drum subspace, the response was correct, and the percentage of correct answers was calculated.

\subsection{Result}
\begin{table}[t]
\centering
\vspace*{1mm}
\scalebox{0.7}{
\begin{tabular}{c|c|c|c|c}
\hline
&\multicolumn{2}{|c|}{Input 3 seconds of data}&\multicolumn{2}{|c}{Input 10 seconds of data}\\
\hline
    instrument &proposed[\%]&separated[\%]&proposed[\%]&separated[\%]\\
 \hline
 drums&82.89&87.90&89.69&88.91\\
 bass&61.94&51.46&84.45&64.87\\
 piano&67.40&47.00&85.70&50.34\\
 guitar&70.15&-&86.26&-\\
 others&74.08&-&84.86&-\\
 \hline
 \end{tabular}
 }
 \caption{kNN-based classification accuracy using each feature representation. The line “proposed” and “separated” show the result using the proposed method and using the separated sounds as input to the individual networks of the conventional method. }
 \label{table:knn}
\end{table}
\subsubsection{Accuracy of embedding}
The accuracy rate of the predicted music IDs is shown in Table \ref{table:knn}. The table shows the results when 3 and 10 seconds of data were used as input for inference.
Each row represents the instrument to focus on. The column for the proposed method shows the results of inference using only the subspace to which the focused instrument sound is assigned. In contrast, the column for the conventional method shows the results of inputting the separated instrument sound to the individual networks. 
\par
\begin{figure}[t]
\begin{minipage}[t]{0.45\columnwidth}
    \centering
    \includegraphics[scale=0.31]{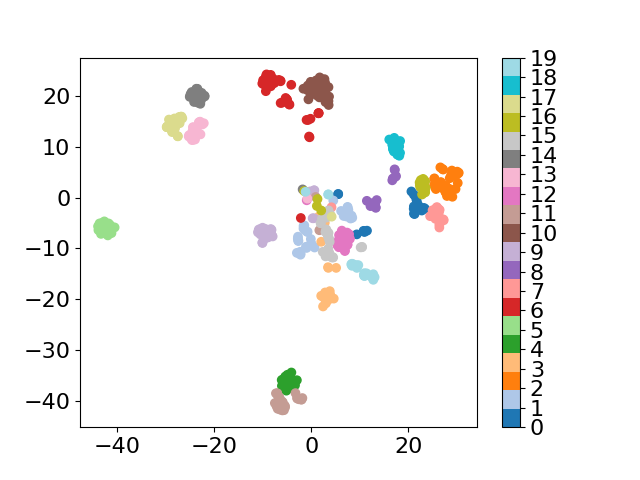}
    \subcaption{The drums' subspace}
    \end{minipage}
  \begin{minipage}[t]{0.45\linewidth}
    \centering
    \includegraphics[scale=0.31]{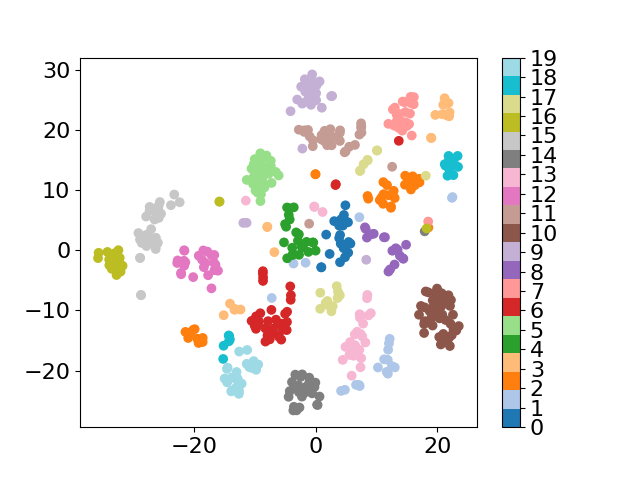}
    \subcaption{The guitar's subspace}
\end{minipage}
\caption{An example of visualized feature representations. We used t-SNE \cite{tsne} to compress the 650-dimensional feature representations to two-dimensional representations. The numbers on the right side of the color bars show the music IDs of 20 pieces from the test set. Segments from the same musical piece are plotted with the same color in both diagrams.}
\label{emb}
\end{figure}

It can be seen that the conventional method using instrumental sound separation is affected by sound quality degradation due to separation, and the accuracy of feature representation degrades, especially on bass and piano with low separation accuracy~\cite{has2022}. In contrast, the proposed method shows stable accuracy regardless of which instrument is focused on.
An example of the subspace is shown in Fig.~\ref{emb}. It can be seen that segments from the same musical piece constitute a cluster and that the subspace can be learned with different distance relationships between the musical pieces.

\subsubsection{Representation of each subspace}
\begin{figure}[t]
\begin{minipage}[t]{0.48\columnwidth}
    \centering
    \includegraphics[scale=0.33]{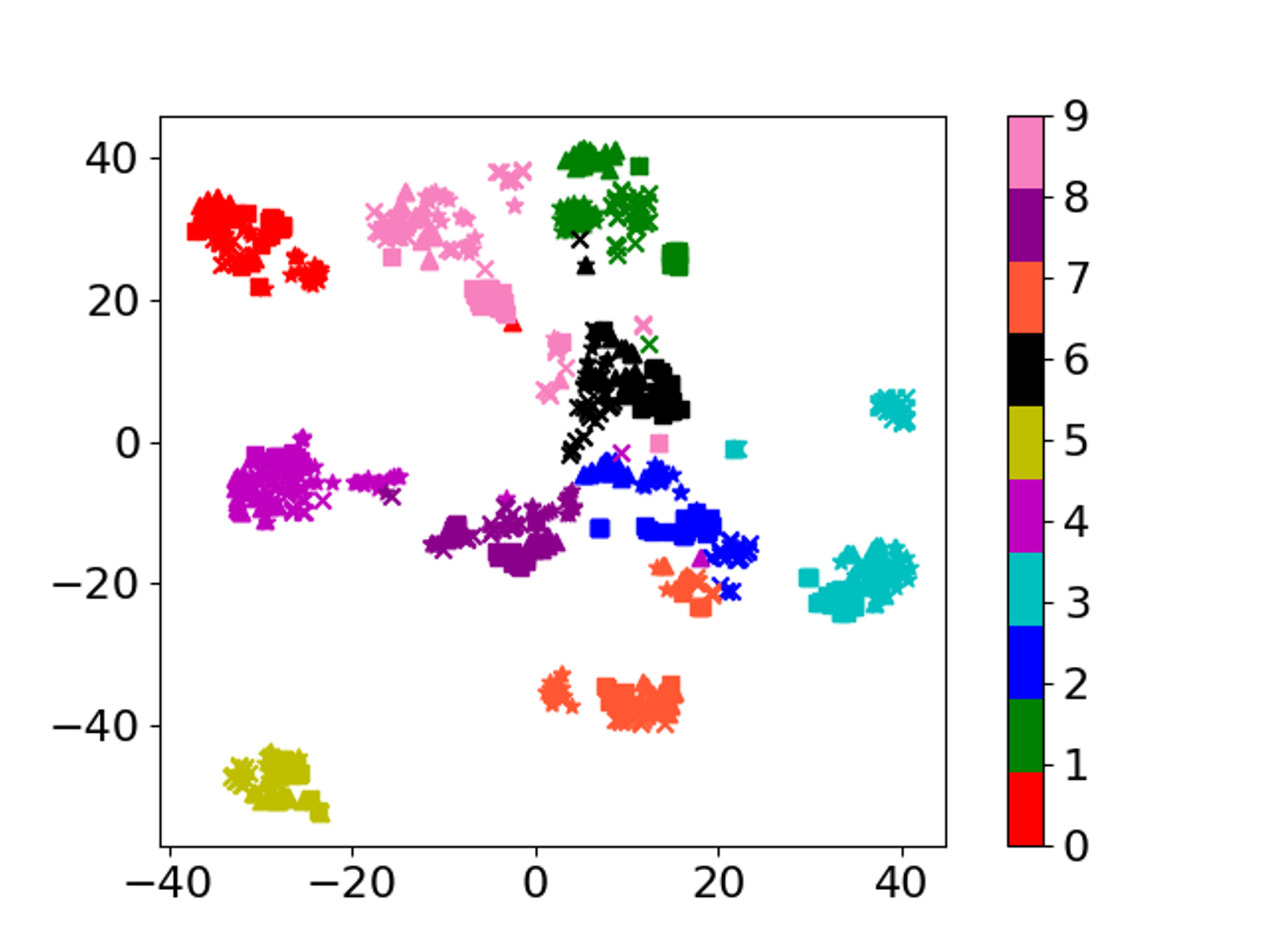}
    \subcaption{The drums' subspace}
    \end{minipage}
  \begin{minipage}[t]{0.48\linewidth}
    \centering
    \includegraphics[scale=0.33]{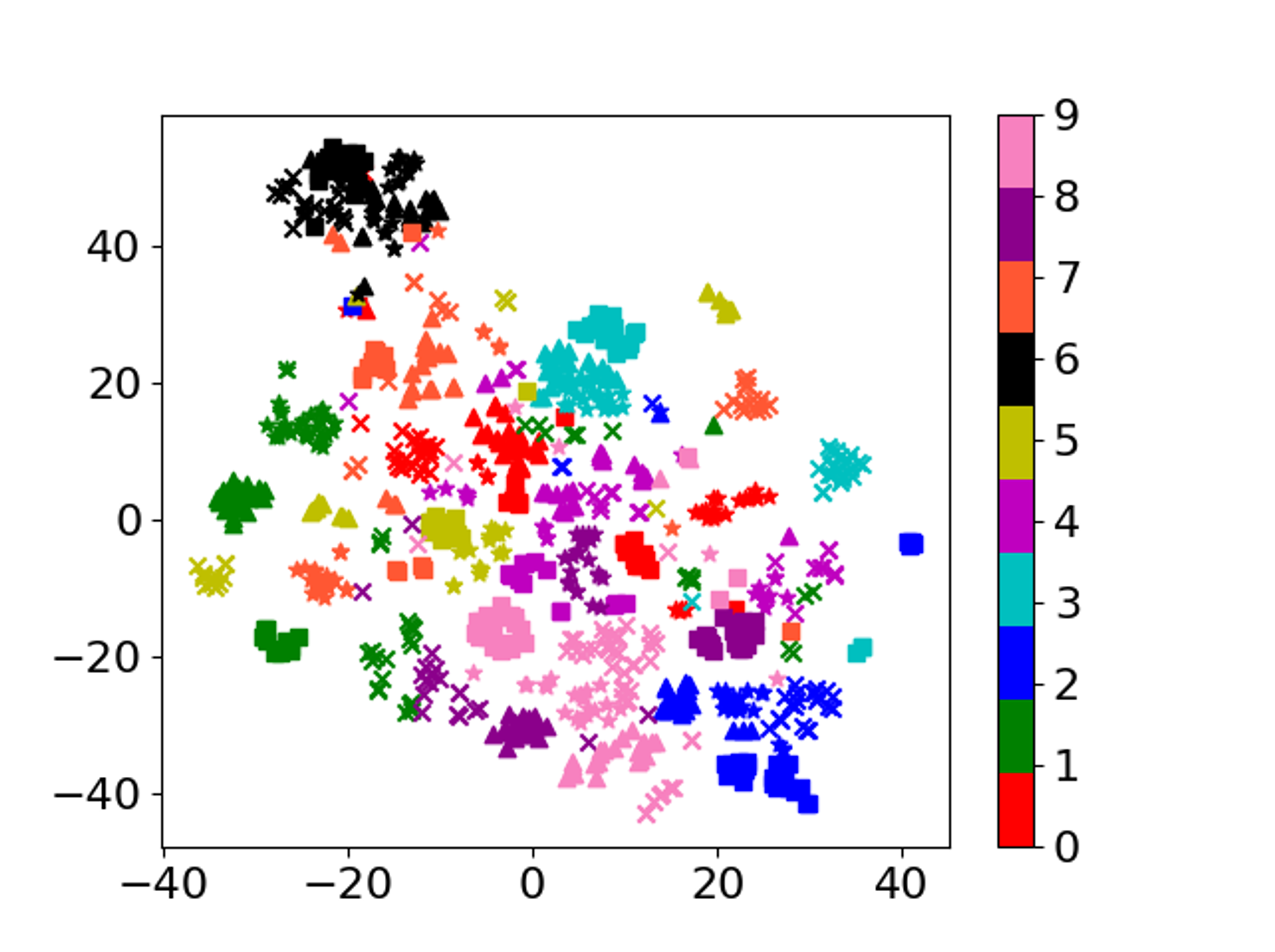}
    \subcaption{The bass's subspace}
\end{minipage}
\caption{An example of visualized feature representations. The numbers on the right side of the color bars show the music IDs of 10 test musical pieces. For example, in (a), segments divided from musical pieces containing the same drum sound were plotted in the same color, and different symbols ($\times$, $\star$, $\blacktriangle$, $\blacksquare$) were used for different instrumental sounds other than the drums contained within those segments. Samples with the same color and the same symbol were not used for reference when evaluating by kNN.}
\label{emb40}
\end{figure}

Table~\ref{table:knnsub} shows the evaluation results for each subspace using pseudo-mixed pieces. Drums and others show high accuracy, indicating that the subspace retains the information of the instrument to which it is assigned. On the other hand, guitars, basses, and pianos show lower values. However, in predicting one label from 10 labels, all the instruments yielded values above 30\%, indicating that each subspace retains some information about the assigned instrument. It also shows that the additional triplet described in Section~\ref{lossfunc} is useful for the semantic separation of spaces. All segments of the 40 pseudo-mixed pieces used in the test, divided into 10-second segments, are plotted and visualized in two dimensions in Fig.~\ref{emb40}. It can be seen that the pieces with the same instrumental sound labels are close to each other.

 \begin{table}[t]
\setlength{\tabcolsep}{4pt}
\centering
\scalebox{0.85}{
\begin{tabular}{c||c|c|c|c|c}
\hline
 instrument&drums&bass&piano&guitar&others\\
 \hline
 basic triplet[\%]&73.1&28.9&26.1&35.7&67.2\\
 aux+basic[\%]&76.8&25.7&23.5&35.9&72.0\\
 aux+basic+add[\%]&82.5&31.8&28.5&40.4&72.5\\
 \hline
aux+basic+add'[\%]&85.5&37.1&31.3&44.7&74.7\\
 \hline
 \end{tabular}
 }
 \caption{kNN-based classification accuracy using the pseudo-mixed piece. The “basic triplet,” “aux+basic,” and “aux+basic+add” mean that model was trained using only the basic triplets, using the basic triplets and the auxiliary loss, and using the basic and additional triplets and the auxiliary loss, respectively. The top three are the results of training with 200 pieces, and the bottom one is the result of training with 1,200 pieces.}
 \label{table:knnsub}
 \end{table}

\subsubsection{Subjective evaluation}


Table~\ref{table:subTR} shows the results of the subjective experiment. The “set 1” and “set 2” denote results of the $\{x, a, b\}$ set and the $\{x, y, c\}$ set, respectively, described in Section~\ref{sub}, and "total" shows the combined results of these two sets. As observed in "total", all cases except for piano were higher than 70\%. Especially, as shown in "set 2", different time segments within the same piece are perceived by humans as similar to each other, and their distances are learned small in the proposed method. In "set 1", as the similarity between different pieces are evaluated, the correct rates tend to degrade owing to a more difficult task. Nevertheless, in the results for drums and guitar, the percentage of correct answers is still high, and this suggests these subspaces well capture the similarity focusing on each instrumental sound.
These results indicated that the music similarity metrics learned with the proposed method focusing on a specific type of sound could find perceptually similar segments in terms of the focused perspective, and they corresponded relatively well to the perception of human senses in some instruments.

\begin{table}[h]
\setlength{\tabcolsep}{4pt}
\centering
\scalebox{0.8}{
\begin{tabular}{c||c|c|c|c}
\hline
 instrument&drums&bass&piano&guitar\\
 \hline
 set 1 [\%]&70.8±8.2&53.3±9.0&44.2±8.9&84.2±6.6\\
 set 2 [\%]&82.5±6.8&98.3±2.3&74.2±7.9&80.8±7.1\\
 total [\%]&76.7±5.4&75.8±5.4&59.2±6.2&82.5±4.8\\
 \hline
 \end{tabular}
 }
 \caption{Correct rates and 95\% confidence intervals as a result of subjective evaluation}
 \label{table:subTR}
 \end{table}

\section{Conclusion}

In this paper, we propose a method to compute similarities focusing on each instrumental sound using mixed sounds as input in one network, which extracts a single similarity embedding space with disentangled dimensions for each instrument. Experimental results have shown each subspace represents a similarity criterion focused on the instrument to which it is assigned especially in drums, guitar, and others. Future work includes application to data sets with vocals.

\clearpage
\bibliography{reference}

\begin{thebibliography}{10}
\providecommand{\url}[1]{#1}
\csname url@samestyle\endcsname
\providecommand{\newblock}{\relax}
\providecommand{\bibinfo}[2]{#2}
\providecommand{\BIBentrySTDinterwordspacing}{\spaceskip=0pt\relax}
\providecommand{\BIBentryALTinterwordstretchfactor}{4}
\providecommand{\BIBentryALTinterwordspacing}{\spaceskip=\fontdimen2\font plus
\BIBentryALTinterwordstretchfactor\fontdimen3\font minus
  \fontdimen4\font\relax}
\providecommand{\BIBforeignlanguage}[2]{{%
\expandafter\ifx\csname l@#1\endcsname\relax
\typeout{** WARNING: IEEEtran.bst: No hyphenation pattern has been}%
\typeout{** loaded for the language `#1'. Using the pattern for}%
\typeout{** the default language instead.}%
\else
\language=\csname l@#1\endcsname
\fi
#2}}
\providecommand{\BIBdecl}{\relax}
\BIBdecl

\bibitem{ifpi}
IFPI, ``Global music report 2022,'' 2022,
  \url{https://www.ifpi.org/wp-content/uploads/2022/04/IFPI_Global_Music_Report_2022-State_of_the_Industry.pdf}.

\bibitem{apple}
{Apple Inc}, ``Apple music,'' 2023,
  \url{https://www.apple.com/jp/apple-music/}.

\bibitem{Hamel2010}
P.~Hamel and D.~Eck, ``Learning features from music audio with deep belief
  networks,'' in \emph{International Society for Music Information Retrieval
  Conference}, 2010, pp. 339--344.

\bibitem{Elbir2020}
A.~Elbir and N.~Aydin, ``Music genre classification and music recommendation by
  using deep learning,'' \emph{Electronics Letters}, vol.~56, no.~12, pp.
  627--629, 2020.

\bibitem{park2018}
J.~Park, J.~Lee, J.~Park, J.~Ha, and J.~Nam, ``Representation learning of music
  using artist labels,'' in \emph{International Society for Music Information
  Retrieval Conference}, 2018, pp. 717--724.

\bibitem{Clevelan2020}
\BIBentryALTinterwordspacing
J.~Cleveland, D.~Cheng, M.~Zhou, T.~Joachims, and D.~Turnbull, ``Content-based
  music similarity with triplet networks,'' 2020. [Online]. Available:
  \url{[https://arxiv.org/abs/2008.04938]}
\BIBentrySTDinterwordspacing

\bibitem{Lu2017}
R.~Lu, K.~Wu, Z.~Duan, and C.~Zhang, ``Deep ranking: Triplet matchnet for music
  metric learning,'' in \emph{IEEE International Conference on Acoustics,
  Speech and Signal Processing}, 2017, pp. 121--125.

\bibitem{has2022}
Y.~Hashizume, L.~Li, and T.~Toda, ``Music similarity calculation of individual
  instrumental sounds using metric learning,'' in \emph{Asia-Pacific Signal and
  Information Processing Association Annual Summit and Conference}, 2022, pp.
  33--38.

\bibitem{has2022au}
Y.~Hashizume, L.~Li, and T.~Toda, ``Evaluation of music similarity learning
  focusing on each instrumental sound,'' in \emph{Proc. of Autumn Meeting of
  ASJ (in Japanese) 3-1-5}, 2022, pp. 1517--1518.

\bibitem{Bengio2013}
Y.~Bengio, ``Deep learning of representations: Looking forward.'' in
  \emph{International Conference on Statistical Language and Speech
  Processing}, 2013, pp. 1--37.

\bibitem{Hsu2018}
\BIBentryALTinterwordspacing
W.-N. Hsu, Y.~Zhang, R.~J. Weiss, H.~Zen, Y.~Wu, Y.~Wang, Y.~Cao, Y.~Jia,
  Z.~Chen, J.~Shen, P.~Nguyen, and R.~Pang, ``Hierarchical generative modeling
  for controllable speech synthesis,'' 2018. [Online]. Available:
  \url{[https://arxiv.org/abs/1810.07217]}
\BIBentrySTDinterwordspacing

\bibitem{Hsu2019}
W.-N. Hsu, Y.~Zhang, R.~J. Weiss, Y.-A. Chung, Y.~Wang, Y.~Wu, and J.~Glass,
  ``Disentangling correlated speaker and noise for speech synthesis via data
  augmentation and adversarial factorization,'' in \emph{IEEE International
  Conference on Acoustics, Speech and Signal Processing}, 2019, pp. 5901--5905.

\bibitem{Hung2018}
\BIBentryALTinterwordspacing
Y.-N. Hung, Y.~Chen, and Y.-H. Yang, ``Learning disentangled representations
  for timber and pitch in music audio,'' 2018. [Online]. Available:
  \url{[https://arxiv.org/abs/1811.03271]}
\BIBentrySTDinterwordspacing

\bibitem{Luo2019}
K.~A. Y-J.~Luo and D.~Herremans, ``Learning disentangled representations of
  timbre and pitch for musical instrument sounds using gaussian mixture
  variational autoencoders,'' in \emph{International Society for Music
  Information Retrieval Conference}, 2019, pp. 746--753.

\bibitem{tanaka21}
K.~Tanaka, R.~Nishikimi, Y.~Bando, K.~Yoshii, and S.~Morishima, ``Pitch-timbre
  disentanglement of musical instrument sounds based on vae-based metric
  learning,'' in \emph{IEEE International Conference on Acoustics, Speech and
  Signal Processing}, 2021, pp. 111--115.

\bibitem{CSN}
A.~Veit, S.~Belongie, and T.~Karaletsos, ``Conditional similarity networks,''
  in \emph{IEEE Conference on Computer Vision and Pattern Recognition}, 2017,
  pp. 1781--1789.

\bibitem{Disen}
J.~Lee, N.~J. Bryan, J.~Salamon, Z.~Jin, and J.~Nam, ``Disentangled
  multidimensional metric learning for music similarity,'' in \emph{IEEE
  International Conference on Acoustics, Speech and Signal Processing}, 2020,
  pp. 6--10.

\bibitem{Ailon2015}
E.~Hoffer and N.~Ailon, ``Deep metric learning using triplet network,'' in
  \emph{Similarity-Based Pattern Recognition}, 2015, pp. 84--92.

\bibitem{slakh}
E.~Manilow, G.~Wichern, P.~Seetharaman, and J.~Le~Roux, ``Cutting music source
  separation some slakh: A dataset to study the impact of training data quality
  and quantity,'' in \emph{IEEE Workshop on Applications of Signal Processing
  to Audio and Acoustics}, 2019, pp. 45--49.

\bibitem{unet}
O.~Ronneberger, P.~Fischer, and T.~Brox, ``U-net: Convolutional networks for
  biomedical image segmentation,'' in \emph{Medical Image Computing and
  Computer-Assisted Intervention}, 2015, pp. 234--241.

\bibitem{singunet}
A.~Jansson, E.~J. Humphrey, N.~Montecchio, R.~M. Bittner, A.~Kumar, and
  T.~Weyde, ``Singing voice separation with deep u-net convolutional
  networks,'' in \emph{International Society for Music Information Retrieval
  Conference}, 2017.

\bibitem{tsne}
L.~van~der Maaten and G.~Hinton, ``Visualizing data using t-sne,''
  \emph{Journal of Machine Learning Research}, vol.~9, no.~86, pp. 2579--2605,
  2008.

\end{thebibliography}

\end{document}